\documentclass[12pt,english]{article}
\usepackage[T1]{fontenc}
\usepackage[latin9]{inputenc}
\usepackage{mathrsfs}
\usepackage{amsmath}
\usepackage{amsthm}
\usepackage{amssymb}
\usepackage{geometry}
\usepackage{setspace}
\usepackage[authoryear]{natbib}
\makeatletter
\theoremstyle{plain}
\newtheorem{prop}{\protect\propositionname}
\theoremstyle{definition}
\newtheorem{defn}{\protect\definitionname}
\theoremstyle{plain}
\newtheorem{thm}{\protect\theoremname}
\theoremstyle{plain}
\newtheorem{assumption}{\protect\assumptionname}
\theoremstyle{plain}
\newtheorem{lem}{\protect\lemmaname}

\@ifundefined{date}{}{\date{}}

\usepackage{nicefrac}
\usepackage{setspace}
\usepackage{chngcntr}
\usepackage{microtype}
\usepackage{needspace}
\usepackage{lmodern}
\usepackage{tikz}
\usepackage{tkz-tab}
\usepackage{caption}
\usepackage{subcaption}
\usepackage{graphicx}

\usepackage{abstract}

\usepackage{parskip}

\title{Identification Design}
\usepackage{amsmath,amsthm,amssymb,amsfonts} 

\author{Maxwell Rosenthal\thanks{Georgia Institute of Technology.  Email: \href{mailto:rosenthal@gatech.edu}{rosenthal@gatech.edu.} I thank Marciano Siniscalchi and two anonymous referees for valuable feedback, including the suggested relabeling of the paper's title and central exercise from \textit{prior-free information design} to \textit{identification design}. This paper has benefited from conversations with Daniel Dench and Mark Whitmeyer.}}

\date{September 10, 2026}

\def\D{\;\mathrm{d}}

\usepackage[colorlinks=true,linkcolor=blue,citecolor=blue,urlcolor=blue,hyperfootnotes=false]{hyperref}

\makeatother

\usepackage{babel}
\providecommand{\assumptionname}{Assumption}
\providecommand{\definitionname}{Definition}
\providecommand{\lemmaname}{Lemma}
\providecommand{\propositionname}{Proposition}
\providecommand{\theoremname}{Theorem}

\begin{document}
\maketitle
\begin{abstract}
\setstretch{1.0}
This paper introduces a model of \textit{identification design} in which the decision maker observes the population distribution of signals generated by an information structure and ranks actions by their worst-case payoff over the admissible state distributions consistent with those signals.  The environment is \textit{manipulable} if every action is implementable under every admissible distribution of the state variable, and \textit{strongly} so if this is achievable via \textit{almost fully informative} information structures that conceal at most one dimension of information from the decision maker.  We show that manipulability holds if and only if every action has the same worst-case payoff if and only if strong manipulability holds.  In our main application to robust causal inference in econometrics, we verify that treatment-effects models satisfy both manipulability criteria.  More broadly, we provide a complete characterization of the set of implementable actions for all non-manipulable environments in which the decision maker's underlying payoffs are of the expected utility form and the set of admissible state distributions is compact and convex.

\end{abstract}



\section{Introduction}

Frequentist data analysis is standard in regulatory evaluations of
medical interventions and in scientific publishing. At the same time,
researchers are themselves economic agents who face private incentives
and may therefore be tempted to selectively disclose information,
even when what they do disclose must accurately reflect their data.
How should an objective decision maker respond?

Existing models of strategic communication typically study Bayesian
decision makers who filter evidence through an exogenous prior. In
our model, the decision maker instead observes the population distribution
of signals generated by an information structure. He understands the
stochastic relationship between states and signals; views any admissible
distribution of states that rationalizes the observed distribution
of signals as plausible; and evaluates actions by their worst-case
payoff over the set of all such state distributions. An information
structure \emph{implements }an action if that action is worst-case
optimal under that structure. 

This framework is directly applicable to causal inference. Consider
the classical treatment-effects environment with binary outcome $Y$,
binary treatment $T$, covariate $X$, and potential outcomes $(Y_{0},Y_{1})$.
Treatment is unconfounded and the assignment mechanism $P(T\mid X)$
assigns each treatment to every covariate group with positive probability.
If the policymaker observes the joint distribution of $(Y,T,X)$ then
the average counterfactual outcomes $\mathbb{E}[Y_{0}]$ and $\mathbb{E}[Y_{1}]$
are exactly identified. In contrast, if the policymaker observes only
the marginal distribution of $(Y,T)$ then the joint distribution
of outcomes, treatments, and covariates is partially identified and
the consequences of treating the untreated are uncertain. As we demonstrate
in Section \ref{Section: motivating example}, there are situations
in which a fully informed policymaker chooses one treatment while
a partially informed policymaker chooses another. 

We develop four sets of formal results. First, because state distribution
$\nu$ is observationally equivalent to the true distribution $\mu$
under information structure $(\Sigma,E)$ if and only if $\nu$ and
$\mu$ generate the same distribution of messages under the stochastic
map $E$, comparisons between information structures $(\Sigma,E),(\Sigma',E')$
reduce to comparisons between null spaces $\ker E,\ker E'$. We completely
characterize the set of all such kernels in Proposition \ref{Prop which null}. 

Second, we call the environment \emph{manipulable }if for every action
$\alpha$ and every admissible state distribution $\mu$ there exists
an information structure that implements $\alpha$ under $\mu$; an
information structure $(\Sigma,E)$ \emph{almost fully informative
}if the null space of $E$ has dimension at most $1$; and the environment
\emph{strongly manipulable }if for every action $\alpha$ and every
admissible state distribution $\mu$ there exists an almost fully
informative information structure that implements $\alpha$ under
$\mu$. In Theorem \ref{thm Manipulability}, we show that the environment
is manipulable if and only if every action has the same worst-case
payoff if and only if the environment is strongly manipulable. Further,
in Proposition \ref{prop: researcher}, we show that every manipulable
environment admits, for each admissible true distribution $\mu$,
a fixed almost fully informative information structure that implements
the researcher's most-preferred action in every counterfactual specification
$\nu$ of the true state distribution that is observationally equivalent
to $\mu$. Even if the policymaker knows the researcher's preferences,
knows that the researcher knows the true distribution of the state
variable, and anticipates a researcher-optimal information structure,
it is impossible for him to refine the identified set by relating
the information structure he is served with to the researcher's goals
and private information.

Third, we apply our characterization of manipulable environments to
general treatment-effects environments with many treatments, many
covariates, and many outcomes. We confirm in Proposition \ref{prop: manip}
that all such environments are strongly manipulable. In the typical
cases in which the true distribution of outcomes under the intended
treatment does not assign certainty to the worst-possible realization
of the outcome variable, we clarify that manipulation is achievable
through disclosures under which (i) the decision maker strictly prefers
that treatment to all other treatments and (ii) the true distribution
of outcomes under that treatment is exactly identified.

Fourth, we return to the general model and characterize the set of
implementable actions for all non-manipulable environments in which
the decision maker's underlying payoffs are of the expected utility
form and the set of admissible state distributions is compact and
convex. In all such cases, implementability is equivalent to implementability
via almost fully informative information structures.

\paragraph*{Institutional context}

The misalignment between researcher incentives and public interest
is an acknowledged phenomenon in the conveyance of scientific evidence,
as discussed in \citet{Spiess2025} and taken as a premise of that
paper and at least one other recent study in econometrics (\citealp{KasySpiess2024}).
While we are distinguished from these works by our emphasis on identification
rather than finite-sample issues, the available evidence suggests
that the targeting of statistical significance is not the only mechanism
for researcher manipulation. Indeed, the sensitivity of coefficient
magnitudes to the choice of control variables is a longstanding concern
in econometrics, as prominently articulated in \citet{Leamer1983}'s
critique of specification search. More recently and more broadly,
\citet{ChristensenMiguel2018} survey the prevalence of both specification
search and other related credibility concerns in applied economics.

In a recent study, \citet{LenzSahn2021} find that over 30\% of sampled
observational studies published in the \emph{American Journal of Political
Science }achieve statistical significance only through the inclusion
of covariates in their regression specifications, without disclosing
or justifying the resulting increases in estimated effect sizes. Lenz
and Sahn emphasize that these inclusions typically increased the magnitude
of the estimated effects rather than the precision with which they
were estimated. Elsewhere, in a systematic review of 22 cohort studies
covering 3,140 randomized clinical trials, \citet{Dwan2014} identify
various types of discrepancies in the reported use of control variables
within and across trial documents in 46--82\% of relevant trials.

Aside from the direct evidence that researchers sometimes engage in
selective reporting practices, there is an abundance of indirect evidence
that this behavior is treated as a serious concern in the form of
institutional guidelines aimed at inhibiting it. The 2007 Food and
Drug Administration Amendments Act (\citealp{FDAAA2007}) mandates
the registration of applicable clinical trials on ClinicalTrials.gov
within 21 days of first enrollment, including a declaration of outcome
measures. More recent FDA guidance on covariate adjustment (\citealp{FDA2023})
recommends prospective specification of covariate adjustment procedures
before unblinding of comparative data; earlier guidance from the International
Conference on Harmonisation (\citealp{ICH1998}) calls for the pre-specification
of statistical analysis plans in pharmaceutical trials. 

On the scientific side, the CONSORT statement for randomized controlled
trials (\citealp{Hopewell2025}) provides guidelines for the reporting
of randomization procedures, access to trial protocols and statistical
analysis plans, and deviations from prespecified outcomes or analyses.
Within economics, the American Economic Association requires field
experiments submitted to its society journals to be registered before
submission (\citealp{AEA2026RCTRegistry}) and allows researchers
to disclose statistical analysis plans (\citealp{Olken2015}); separately,
the \emph{Journal of Development Economics} offers an editorial process
in which authors disclose their full statistical analysis plan and
the journal commits in principle to publication before the results
are known (\citealp{BogdanoskiFosterKarlanMiguel2018}). As with regulatory
guidelines, all of these practices either constrain the researcher's
choice of information structure or make deviations from declared information
structures observable ex post. The value of such commitment devices
is illustrated by \citet{CaseyGlennersterMiguel2012}, who demonstrate
that their preregistered analysis plan ruled out analyses that would
have variously yielded both positive and negative effects of the evaluated
program on local institutions.

\paragraph*{Roadmap}

The paper is organized as follows. We illustrate the mechanics of
our framework in Section \ref{Section: motivating example}, lay out
the model in Section \ref{SECTION: MODEL}, and discuss related literature
in Section \ref{section: related literature}. We characterize manipulable
environments in Section \ref{Section: Manip}, apply the framework
to treatment-effects models in Section \ref{Section: TE}, and characterize
implementability in non-manipulable environments in Section \ref{sec:Non-manipulable-environments}.
Section \ref{section: conclusion} concludes. Proofs and omitted supporting
results are organized into Appendices \ref{appendix: manip}--\ref{appendix: non manip}.
Appendix \ref{AppendX: regret} develops a partial extension of our
results under the alternative minimax regret decision making criterion.

\section{\label{Section: motivating example}Motivating Example}

A researcher discloses data from an observational study of treatment
outcomes to a policymaker. The treatment $T$, untreated outcome $Y_{0}$,
treated outcome $Y_{1}$, and covariate $X$ are each binary. The
researcher observes the true joint distribution $\mu$ of $(Y,T,X)$
shown in Table \ref{t1}; the observed outcome $Y$ satisfies the
\emph{consistency }assumption $Y=Y_{T}$; the counterfactual outcomes
$(Y_{0},Y_{1})$ satisfy the \emph{unconfoundedness }assumption $(Y_{0},Y_{1})\perp T\mid X$;
and the assignment mechanism $\mu(T\mid X)$ satisfies the \emph{strict
overlap }assumption $0<\mu(T=t\mid X=x)<1$ for all $t,x$.

\begin{table}[h!]
\centering
\caption{Observed distribution $\mu(Y,T,X)$}
\setlength{\tabcolsep}{10pt}
\renewcommand{\arraystretch}{1.2}
\label{t1}
\begin{tabular}{c|cccc}
  & $(T{=}0,X{=}0)$ & $(T{=}0,X{=}1)$ & $(T{=}1,X{=}0)$ & $(T{=}1,X{=}1)$ \\
\hline
$Y=0$ & 0.40 & 0.05 & 0.10 & 0.30 \\
$Y=1$ & 0.00 & 0.05 & 0.00 & 0.10 \\
\end{tabular}
\end{table}
The policymaker's problem is to choose the treatment $a\in\{0,1\}$
that robustly maximizes the average outcome $\mathbb{E}[Y_{a}]$.
If the researcher discloses the full joint distribution $\mu$, then
the counterfactual means
\begin{align}
\mathbb{E}[Y_{0}] & =\mathbb{E}[\mathbb{E}[Y\mid T=0,X]]=\sum_{y,x}\frac{y\cdot\mu(y,0,x)}{\mu(T=0\mid X=x)}=0.25,\label{eq1}\\
\mathbb{E}[Y_{1}] & =\mathbb{E}[\mathbb{E}[Y\mid T=1,X]]=\sum_{y,x}\frac{y\cdot\mu(y,1,x)}{\mu(T=1\mid X=x)}=0.125\label{eq2}
\end{align}
are exactly identified and the policymaker declines to treat. If instead
the researcher discloses only the marginal distribution of $(Y,T)$,
then the policymaker evaluates treatment $a$ by the worst-case instantiation
of $\mathbb{E}_{\nu}[Y_{a}]$ against all joint distributions $\nu$
on $(Y,T,X)$ consistent with the disclosed marginal in Table \ref{t2}
and our maintained unconfoundedness and strict overlap assumptions. 

\begin{table}[h!]
\centering
\caption{Disclosed marginal distribution $\mu(Y,T)$}
\setlength{\tabcolsep}{15pt}
\renewcommand{\arraystretch}{1.2}
\label{t2}
\begin{tabular}{c|cc}
  & $T=0$ & $T=1$ \\
\hline
$Y=0$ & 0.45 & 0.40 \\
$Y=1$ & 0.05 & 0.10 \\
\end{tabular}
\end{table}

 As the inverse-probability weighting formulae (\ref{eq1})--(\ref{eq2})
make clear, the worst case for action $a$ is that treatment $T=a$
was assigned as frequently as possible to $X$-groups that benefit
most from it. While the disclosed marginal distribution of $(Y,T)$
constrains these quantities, it does not identify them. The limiting\footnote{The inverse probability weight formulae (\ref{eq1})--(\ref{eq2})
are well defined as long as the maintained strict overlap assumption
is satisfied. Accordingly, the joint distribution in the third table
should be interpreted formally as the limit of a particular sequence
of joint distributions each with that property.} distribution $\nu$ in Table \ref{t3} is consistent with the disclosed
marginal and yields worst-case payoffs
\begin{align*}
\mathbb{E}_{\nu^ {}}[Y_{0}] & =0.05, & \mathbb{E}_{\nu^ {}}[Y_{1}] & =0.10
\end{align*}
for both treatments. Although treatment is suboptimal when the policymaker
observes the full joint distribution of $(Y,T,X)$, it is robustly
optimal when he observes only the marginal of $(Y,T)$. Uncertainty
about observation-level treatment propensities reverses the full-information
optimal policy.

\begin{table}[h!]
\centering
\caption{Limiting worst-case distribution $\nu$ of $(Y, T, X)$ under partial disclosure}
\vspace{0.5em}

\setlength{\tabcolsep}{10pt}
\renewcommand{\arraystretch}{1.2}
\label{t3}
\begin{tabular}{c|cccc}
  & $(T{=}0,X{=}0)$ & $(T{=}0,X{=}1)$ & $(T{=}1,X{=}0)$ & $(T{=}1,X{=}1)$ \\
\hline
$Y=0$ & 0.45 & 0 & 0 & 0.40 \\
$Y=1$ & 0.05 & 0 & 0 & 0.10 \\
\end{tabular}
\end{table}

\section{\label{SECTION: MODEL}Model}

We write $\mathbb{R}^{n}$ for the set of real vectors of length $n$
equipped with the standard metric and identify real-valued functions
on finite sets $X$ with vectors in $\mathbb{R}^{\vert X\vert}$.
Given a finite set $X$, we write $\Delta(X)$ for the set of all
probabilities on $X$ and interpret the elements of $\Delta(X)$ as
real vectors. More broadly, given a metric space $X$, we give the
set of Borel distributions $\Delta(X)$ the topology of weak convergence
and write $\text{supp}(\nu)$ for the support of $\nu\in\Delta(X)$.

This paper makes use of standard results from linear algebra. We write
$\text{span}(S)$ for the span of a set of real vectors $S\subset\mathbb{R}^{n}$
and $\text{span}\{d\}$ for the span of a single vector $d\in\mathbb{R}^{n}$.
More generally, if $S$ is a subset of a vector space we write $\text{dim}(S)$
for the dimension of its affine hull. Finally, given a linear map
$f:\mathbb{R}^{n}\to\mathbb{R}^{m}$, we write $\ker f\equiv\{v\in\mathbb{R}^{n}\mid f(v)=0\}$
for its null space.

\paragraph{States, actions, and beliefs}

The set $\Omega$ of states is non-empty and finite; the set $\mathcal{P}\subset\Delta(\Omega)$
of \emph{admissible }state distributions\footnote{Despite our non-Bayesian framework, we follow the literature on decision
making under uncertainty and occasionally refer to elements of $\mathcal{P}$
as \emph{priors.}} contains the true distribution of the state of the world $\mu$;
and the set $A$ of actions is a non-empty compact metric space. The
decision maker's utility $U:\Delta(A)\times\mathcal{P}\to\mathbb{R}$
is continuous and satisfies 
\[
U(\lambda\alpha+(1-\lambda)\beta,\nu)=\lambda U(\alpha,\nu)+(1-\lambda)U(\beta,\nu)
\]
for all actions $\alpha,\beta$, all admissible priors $\nu$, and
all $\lambda\in[0,1]$. With an eye towards our primary application,
we assume neither that $U$ is affine in its second argument nor that
$\mathcal{P}$ is compact or convex. At the same time, we do require
that the image of $\mathcal{P}$ under $U(\alpha,\cdot)$ is compact
for all actions $\alpha$. 

We write $\omega$ for generic elements of $\Omega$, $a$ for generic
elements of $A$, $\nu$ for generic elements of $\mathcal{P}$, and
$\alpha$ for generic elements of $\Delta(A)$. In an abuse of notation,
we write $U(a,\nu)$ for the decision maker's payoff under the degenerate
mixed action that assigns probability one to $a$. Tuple $(\Omega,\mathcal{P},A,U)$
is the \emph{environment. }

\paragraph{Information structures}

The decision maker acts on objective information about the true distribution
of the state of the world $\mu$. He observes the distribution of
messages generated by \emph{information structure $(\Sigma,E)$, }where
$\Sigma$ is a finite set of messages and $E:\Omega\to\Delta(\Sigma)$
assigns message distributions $E(\cdot\mid\omega)$ to states $\omega$.
Admissible state distribution $\nu$ is \emph{observationally equivalent}
to $\mu$ if and only if $E\nu=E\mu$, and we write 
\begin{equation}
\mathcal{P}_{\mu}(\Sigma,E)\equiv\{\nu\in\mathcal{P}\mid E\nu=E\mu\}=\{\nu\in\mathcal{P}\mid(\nu-\mu)\in\ker E\}\label{display: obs eq}
\end{equation}
for the \emph{identified set }of all such distributions. Information
structure $(\Sigma,E)$ is \emph{fully informative }if $\ker E$ has
dimension $0$; \emph{uninformative }if $\ker E$ has dimension $\vert\Omega\vert-1$;
and \emph{almost fully informative} if $\ker E$ has dimension at
most $1$. We emphasize that fully informative information structures
are almost fully informative and furthermore that they exactly identify
the true state distribution $\mu$, as in the full-disclosure benchmark
of Section \ref{Section: motivating example}.

\paragraph{The decision maker's problem}

The \emph{decision maker's problem }
\begin{equation}
\max_{\alpha\in\Delta(A)}\;\inf_{\nu\in\mathcal{P_{\mu}}(\Sigma,E)}\;U(\alpha,\nu)\label{DM}
\end{equation}
is to maximize his worst-case payoff over the identified set. Information
structure $(\Sigma,E)$ \emph{implements }action $\alpha$ if $\alpha$
is a solution to problem (\ref{DM}) under $(\Sigma,E)$, and \emph{strictly}
so if that solution is unique.

\paragraph{The researcher's problem}

Our model admits two interpretations. First, we imagine settings in
which the information structure is an ambient feature of the environment
that arises exogenously from one or more sources of information available
to the decision maker. Second, we imagine settings in which the information
structure is intentionally designed by a researcher whose goals need
not be aligned with the interests of the decision maker. In the latter,
the researcher observes the true state distribution $\mu$ and her
utility $v:\Delta(A)\to\mathbb{R}$ is upper semicontinuous.

 The \emph{researcher's problem} is then
\[
\max_{(\alpha,\Sigma,E)}\;v(\alpha)\quad\text{subject to}\quad\forall\beta\in\Delta(A)\;\inf_{\nu\in\mathcal{P}_{\mu}(\Sigma,E)}U(\alpha,\nu)\geq\inf_{\nu\in\mathcal{P}_{\mu}(\Sigma,E)}U(\beta,\nu),
\]
in which she chooses an action $\alpha$ and information structure
$(\Sigma,E)$ to maximize $v(\alpha)$ subject to the constraint that
$(\Sigma,E)$ implements $\alpha$. We implicitly resolve ties in
the researcher's favor, and write $\alpha^{FB}$ for an arbitrarily
selected unconstrained maximizer of her utility $v$. 

\section{\label{section: related literature}Related literature}

This paper endogenizes partial identification in the same way that
Bayesian Persuasion (\citealp{KamenicaGentzkow2011}) endogenizes
Blackwell experiments.

Blackwell's seminal papers \citeyearpar{Blackwell1951,Blackwell1953}
introduce a model of statistical experimentation in which a Bayesian
decision maker observes a signal generated by the experiment and updates
his beliefs before acting. Kamenica and Gentzkow endogenize the choice
of experiment in a sender-receiver environment, characterize the sender-optimal
experiment, and thereby launch the Bayesian persuasion literature.
In contrast, our decision maker responds to the entire distribution
of signals rather than to individual realizations, has no prior, and
evaluates actions by their worst-case payoff over the identified set.
While \citet{LinLiu2024} provide a role for the unconditional message
distribution in enforcing the commitment assumption in Bayesian persuasion,
their receiver responds to individual signal realizations rather than
the entire distribution, as in the standard formulation of the persuasion
problem. 

A growing literature studies persuasion and communication with ambiguity-averse
agents. \citet{BeaucheneLiLi2019} study a sender who strategically
deploys ambiguous communication devices to an ambiguity-averse receiver
who starts from a unique prior; \citet{HedlundKauffeldtLammert2021}
study persuasion of a receiver with an exogenous interval of priors
and $\alpha$-maxmin preferences. In both papers, the receiver observes
and responds to individual signal realizations. In a study more closely
related to our own, \citet{KoesslerPahlke2025} design the coarseness
of aggregate feedback about opponents' play in games with ambiguity-averse
players, with beliefs restricted to be consistent with the observed
aggregate data. Their decision makers respond to distributional information
rather than to individual signals, as ours does; we differ in that
our uncertainty is about the payoff-relevant state rather than about
opponents' strategies, and in our consideration of a single-agent
decision problem rather than a game.

Our model of individual decision making was developed in earlier work
(\citealp{Rosenthal2026PFB}). Outside this framework, there is a
small but growing literature on non-Bayesian decision making in Blackwell
experiments and sender-receiver games. \citet{Whitmeyer2026} studies
the information-monotonicity of non-Bayesian updating rules, and \citet{YangYoderZentefis2025}
study the value of finite-dimensional explanations of complex models,
showing that no explanation can improve the payoff of a worst-case
decision maker. An earlier stream of papers extends Blackwell\textquoteright s
classical framework to ambiguity-averse decision makers with exogenously
specified beliefs (\citealp{Celen2012MEU,HeyenWiesenfarth2015MEU,LiZhou2016Blackwell}).

More distantly related is the robust mechanism design (\citealp{BergemannMorris2005})
literature in which the designer is uncertain about various features
of her environment and seeks a mechanism that performs well in all
plausible cases. Our model is most closely related to the branch of
this literature that is concerned with distributional robustness,
wherein the designer knows only moments or other coarse features of
the distribution of the uncertain variables and maximizes her worst-case
payoff against the set of all distributions consistent with those
restrictions. Applications include screening (\citealp{CarrascoEtAl2018}),
auctions (\citealp{BrooksDuFeffer2025}), and contracting (\citealp{Rosenthal2026SIDB}).
While the restrictions on the unknown distribution in those papers
are primitives that constrain the designer, the researcher in our
model creates the restrictions by choosing what to disclose, and it
is the decision maker who responds to them with his worst-case criterion.

The partial identification literature in econometrics initiated by
\citet{Manski1990,Manski1997,Manski2003}
 and surveyed by \citet{Tamer2010}, \citet{Molinari2020}, and \citet{KlineTamer2023}
studies contexts in which economic data are insufficient to exactly
identify the true distribution of relevant variables under the assumptions
the analyst is willing to make about the environment. Our decision-theoretic
framework --- in which the distribution of the state of the world
is set-identified and the decision maker ranks actions according to
their worst-case payoff over the identified set --- is most closely
related to Manski's work on statistical treatment rules (\citealp{Manski2007,Manski2013}).
Aside from our use of the maxmin decision making criterion in our
main specification rather than the minimax regret criterion we treat
in Appendix \ref{AppendX: regret}, we are conceptually distinguished
from the statistical decision theory literature by our endogenization
of the identified set via our interpretation of the information structure
as a choice variable. This facilitates our characterization of implementable
actions and our identification of the almost fully informative information
structures that implement them. 

Less closely related is a recent study by \citet{AndrewsShapiro2021}
that considers the problem faced by a benevolent analyst in reporting
data to an audience of Bayesian receivers with heterogeneous priors.
Elsewhere, \citet{BanerjeeChassangMonteroSnowberg2020} study experiment
design for an ambiguity-averse experimenter who is uncertain about
the distribution of counterfactual outcomes across covariate groups,
with an emphasis on the value of randomized assignment. In contrast
to the motivating premise for our paper, preference misalignment plays
no role in either model.

\section{\label{Section: Manip}Manipulability}

The identified set $\mathcal{P_{\mu}}(\Sigma,E)$ is jointly characterized
by the exogenous set of admissible priors $\mathcal{P}$, the exogenous
true distribution $\mu$, and the endogenous information structure
$(\Sigma,E)$. As in (\ref{display: obs eq}), admissible prior $\nu$
is observationally equivalent to the true distribution $\mu$ if and
only if $\nu-\mu\in\ker E$. From that point of view, our problem
reduces to one in which the independent variable of interest is the
null space of the stochastic map $E$, rather than $(\Sigma,E)$ itself.
\begin{prop}
\label{Prop which null}There exists an information structure $(\Sigma,E)$
with $\ker E=D$ if and only if $D$ is a linear subspace of $\mathbb{R}^{\vert\Omega\vert}$
satisfying $\sum_{\omega\in\Omega}d(\omega)=0$ for all directions
$d\in D$. 
\end{prop}
Proposition \ref{Prop which null} reduces the implementation problem
to a search over kernels. In this section, we study environments in
which that search succeeds for every action, regardless of the true
distribution of the state variable.
\begin{defn}
Environment $(\Omega,\mathcal{P},A,U)$ is \emph{manipulable }if for
all actions $\alpha\in\Delta(A)$ and for all admissible state distributions
$\mu$ there exists an information structure $(\Sigma,E)$ that implements
action $\alpha$ when the true state distribution is $\mu$.
\end{defn}
Manipulability is a demanding criterion. As we show, it holds if and
only if the worst-case payoff for every action is the same.
\begin{defn}
Environment $(\Omega,\mathcal{P},A,U)$ is \emph{strongly manipulable
}if for all actions $\alpha\in\Delta(A)$ and for all admissible state
distributions $\mu$ there exists an almost fully informative information
structure $(\Sigma,E)$ that implements action $\alpha$ when the
true state distribution is $\mu$.
\end{defn}
In addition to establishing the aforementioned characterization of
manipulability, our first theorem clarifies that manipulation is feasible
via the disclosure of at least $\vert\Omega\vert-2$ linearly independent
statistics out of a maximum $\vert\Omega\vert-1$. 

By way of interpretation, in an earlier paper \citep{Rosenthal2026PFB},
we (i) define $(\Sigma,E)$ as \emph{robustly more informative }than
$(\Sigma',E')$ if $\ker E\subset\ker E'$ and (ii) show that one
information structure is robustly more informative than another if
and only if for every decision problem the decision maker's guaranteed
payoff after observing the former matches or exceeds his guaranteed
payoff after observing the latter. While this order provides the ``correct''
notion of relative informativeness in our environment and is consistent
with Blackwell informativeness for any pair of Blackwell-comparable
information structures, there exist strictly Blackwell-comparable
pairs of information structures with identical kernels. Accordingly,
while information structures that conceal exactly one dimension of
information from the decision maker are maximally robustly informative
among the class of information structures that are not fully informative,
they need not be maximally Blackwell informative within the same class.
Any such Blackwell improvement nevertheless generates the same identified
set and thus the same worst-case decision maker payoff for every action.

\begin{thm}
\label{thm Manipulability}The following three statements are equivalent:
\begin{enumerate}
\item[(i)]  Environment $(\Omega,\mathcal{P},A,U)$ is manipulable.
\item[(ii)]  The decision maker's utility satisfies $\min_{\nu\in\mathcal{P}}U(\alpha,\nu)=\min_{\nu\in\mathcal{P}}U(\beta,\nu)$
for every pair of actions $\alpha,\beta$.
\item[(iii)]  Environment $(\Omega,\mathcal{P},A,U)$ is strongly manipulable.
\end{enumerate}
\end{thm}
In Theorem \ref{thm Manipulability}, we identify a straightforward
condition under which the true distribution $\mu$ of the state variable
imposes no constraints on the set of implementable actions. Because
the theorem does not assert the stronger result that there is a fixed
almost fully informative information structure that implements the
desired action across every admissible specification of $\mu$, this
raises the question as to whether or not the decision maker might
be able to make inferences about $\mu$ by attempting to relate the
researcher's disclosure to the researcher's private information. As
we show, there exist researcher-optimal disclosures under which refinement
of the identified set on the basis of the researcher's incentives
is impossible.
\begin{prop}
\label{prop: researcher}If information structure $(\Sigma,E)$ implements
action $\alpha$ under true state distribution $\mu$ then information
structure $(\Sigma,E)$ implements action $\alpha$ under true state
distribution $\nu$ for every $\nu$ in the identified set $\mathcal{P}_{\mu}(\Sigma,E)$.
Consequently, if the environment $(\Omega,\mathcal{P},A,U)$ is manipulable
and the true state distribution is $\mu$ then there exists an almost
fully informative information structure $(\Sigma,E)$ that implements
the first-best action $\alpha^{FB}$ under true distribution $\nu$
for every $\nu$ in the identified set $\mathcal{P}_{\mu}(\Sigma,E)$.
\end{prop}
The first part of Proposition \ref{prop: researcher} follows naturally
from the transitivity of the observational equivalence relation under
any fixed information structure $(\Sigma,E)$, while the second part
is a corollary of the first part and Theorem \ref{thm Manipulability}.
The proposition clarifies that even if the researcher's preferences
and her knowledge of the true state distribution are common knowledge,
there exists a fixed almost fully informative information structure
that implements the first-best action and is consistent with researcher-optimal
play for every state distribution observationally equivalent to the
truth. Under that information structure, no information beyond the
disclosed message distribution is transmitted by the researcher's
choice of disclosure.

\section{\label{Section: TE}Treatment effects}

There are two or more treatments $T_{1},...,T_{k}$ in finite set
$\mathcal{T}$; two or more outcomes $Y$ in finite set $\mathcal{Y}\subset\mathbb{R}$;
and finitely many covariates $X\equiv(X_{1},...,X_{\ell})$. Each
variable $X_{j}$ takes values in non-empty and finite set $\mathcal{X}_{j}$,
and $X$ takes values in $\mathcal{X}\equiv\mathcal{X}_{1}\times...\times\mathcal{X}_{\ell}$.\footnote{Models with only an outcome variable $Y$ and treatment variable $T$
are equivalent to specifications of our model with $\vert\mathcal{X}\vert=1$. } While the latent state $(Y_{1},...,Y_{k},T,X)$ specifies the outcome
$Y_{t}$ under each treatment $t$, the observed state $\omega\equiv(Y_{T},T,X)$
records only the outcome $Y_{T}$ of the treatment actually received
$T$. Unconfoundedness $(Y_{1},...,Y_{k})\perp T\mid X$, together
with strict overlap, identifies the counterfactual mean $\mathbb{E}_{\nu}[Y_{a}]$
from the distribution of observable variables $\nu$, and the set
of actions $A$, the state space $\Omega$, and the decision maker's
utility $U:\Delta(A)\times\mathcal{P}\to\mathbb{R}$ are respectively
\begin{align}
A & \equiv\mathcal{T}, & \Omega & \equiv\mathcal{Y}\times\mathcal{T}\times\mathcal{X}, & U(\alpha,\nu)\equiv\int_{A}\mathbb{E}_{\nu}[Y_{a}]\D\alpha & =\int_{A}\sum_{y,x}\frac{y}{\nu(a\mid x)}\nu(y,a,x)\D\alpha.\label{TE 1}
\end{align}
The assignment mechanism\emph{ }$P(T\mid X)$ belongs to a non-empty
set $\mathscr{P}$ of maps from $\mathcal{X}$ into $\Delta(\mathcal{T})$
satisfying the strict overlap assumption
\begin{equation}
\forall P\in\mathscr{P}\;\forall t\in\mathcal{T}\;\forall x\in\mathcal{X}\;0<P(T=t\mid X=x)<1,\label{TE 2}
\end{equation}
and the set of admissible distributions 
\begin{equation}
\mathcal{P}\equiv\bigcup_{P\in\mathscr{P}}\{\nu\in\Delta(\Omega)\mid\forall t\in\mathcal{T}\;\forall x\in\mathcal{X}\;\nu(t\mid x)=P(t\mid x)\}\label{TE 3}
\end{equation}
is the set of all distributions $\nu$ consistent with the structural
assumptions embedded in $\mathscr{P}$.\footnote{We impose the matching-assignment-mechanism condition $\nu(t\mid x)=P(t\mid x)$
only where $\nu(x)>0$. Zero-probability covariate groups contribute
zero to the payoff sums.} We call environment $(\Omega,\mathcal{P},A,U)$ a \emph{treatment-effects
model }if it satisfies (\ref{TE 1})--(\ref{TE 3}).

In all treatment-effects models $(\Omega,\mathcal{P},A,U)$, the decision
maker's payoffs have a cubical structure under which manipulability
is evident. Prior $\nu$ is admissible if and only if there exists
a family of conditional outcome distributions $\Phi:\mathcal{T}\times\mathcal{X}\to\Delta(\mathcal{Y})$,
an admissible assignment mechanism $P$, and a distribution of weights
$\lambda\in\Delta(\mathcal{X})$ such that 
\begin{align*}
\nu(y,t,x) & =\Phi(y\mid t,x)P(t\mid x)\lambda(x).
\end{align*}
Direct substitution into (\ref{TE 1}) yields
\[
U(a,\nu)=\sum_{y,x}y\Phi(y\mid a,x)\lambda(x).
\]
The conditional distributions $\Phi$ and the weights $\lambda$ are
free parameters; the conditional mean $\sum_{y}y\Phi(y\mid a,x)$
lies anywhere in the interval $[\min\mathcal{Y},\max\mathcal{Y}]$
for each group $(a,x)$; and the set of admissible payoff vectors
is
\begin{equation}
\{(U(a,\nu))_{a\in A}\mid\nu\in\mathcal{P}\}=[\min\mathcal{Y},\max\mathcal{Y}]^{A}.\label{displasy: te cube}
\end{equation}
In turn, treatment-effects models satisfy our maintained assumptions
and the equal-payoff condition of Theorem \ref{thm Manipulability}.
The theorem immediately yields the following result.
\begin{prop}
\label{prop: manip}Treatment-effects models $(\Omega,\mathcal{P},A,U)$
are strongly manipulable.
\end{prop}
While the formal proof of Proposition \ref{prop: manip} follows the
template above, we offer here an alternative constructive argument
that highlights a pair of additional features of our implementing
information structures. Given admissible true distribution $\mu$
and pure action $a$, define 
\[
\nu(y,t,x)\equiv\begin{cases}
\mu(y,t,x) & t=a\\
\mathbf{1}\{y=\min\mathcal{Y}\}\mu(t,x) & t\neq a
\end{cases}
\]
and note that $\nu$ is admissible because $\nu(t,x)=\mu(t,x)$ for
all $(t,x)$ and thus the two priors share an assignment mechanism. 

Consider any almost fully informative information structure $(\Sigma,E)$
with $\ker E=\text{span}\{\nu-\mu\}$. Because $\nu(y,a,x)=\mu(y,a,x)$
for all $(y,x)$, it is straightforward to verify that $\nu'(y,a,x)=\mu(y,a,x)$
for all $(y,x)$ and for all $\nu'$ observationally equivalent to
$\mu$. Further, because $\nu$ is observationally equivalent to $\mu$
and $U(b,\nu)=\min\mathcal{Y}$ for all $b\neq a$, the worst-case
payoff for every distinct action $\alpha$ is no higher than 
\[
U(\alpha,\nu)=\sum_{b\neq a}\alpha(b)\min\mathcal{Y}+\alpha(a)U(a,\mu).
\]
Information structure $(\Sigma,E)$ implements $a$, and strictly
so if $U(a,\mu)$ exceeds $\min\mathcal{Y}$. Moreover, $(\Sigma,E)$
exactly identifies not only the decision maker's payoff under the
intended action $a$, but also the true distribution of outcomes under
$a$.

Because every prior in the identified set shares the true assignment
mechanism, the construction remains valid when that mechanism is known
and when its assignment probabilities are bounded away from zero.

We illustrate the construction by returning to the example in Section
\ref{Section: motivating example}. As demonstrated, disclosing only
the marginal distribution of $(Y,T)$ implements the researcher's
preferred treatment $T=1$ while full disclosure does not. Suppose
instead that the researcher discloses $(Y,T,X)$ except that the two
outcomes in group $(T,X)=(0,1)$ generate the same message. This information
structure is almost fully informative. It exactly identifies the payoff
$0.125$ under treatment $1$, while the worst-case payoff under treatment
$0$ is zero. Accordingly, treatment $1$ is strictly optimal.

\section{\label{sec:Non-manipulable-environments}Non-manipulable environments}

While our treatment-effects application demonstrates that there are
important settings in which Theorem \ref{thm Manipulability}'s equal-floor
characterization of manipulable environments holds, that condition
is not universal. Consider, for instance, a modified version of that
application in which there are heterogeneous and state-independent
costs to treat. Under a set of mild regularity criteria that are ubiquitous
in economic theory, the set of implementable actions in environments
that fail the floor criterion has a straightforward characterization.
\begin{assumption}
\label{Assumption: prior affinity}The decision maker's utility satisfies
\[
U(\alpha,\lambda\nu+(1-\lambda)\nu')=\lambda U(\alpha,\nu)+(1-\lambda)U(\alpha,\nu')
\]
for every action $\alpha$, every pair of admissible priors $\nu,\nu'$,
and every constant $0\leq\lambda\leq1$. 
\end{assumption}
Taken in tandem with our maintained assumption that the decision maker's
utility function $U$ is affine in its first argument, Assumption
\ref{Assumption: prior affinity} implies that the decision maker's
underlying payoffs are of the expected utility form. 
\begin{assumption}
\label{Assumption: compact convex priors}The set of admissible priors
$\mathcal{P}$ is compact and convex.
\end{assumption}
Along with the technical criteria in Assumption \ref{Assumption: compact convex priors},
affinity of the decision maker's underlying payoff function in each
argument is sufficient to imply that the decision maker's problem
given true distribution $\mu$ and information structure $(\Sigma,E)$
has a \emph{saddle point }$(\alpha^{*},\nu^{*})$ satisfying 
\begin{equation}
\forall\alpha\in\Delta(A)\;\forall\nu\in\mathcal{P}_{\mu}(\Sigma,E)\;U(\alpha^{*},\nu)\geq U(\alpha^{*},\nu^{*})\geq U(\alpha,\nu^{*}).\label{display: saddle inequalities}
\end{equation}
Our conditions for implementability build on the saddle inequalities
(\ref{display: saddle inequalities}).\footnote{As observed by \citet{Kuzmics2017}, Wald's complete class theorem
implies that a maxmin decision maker who can randomize over acts chooses
as if he were a subjective expected utility maximizer. We thank an
anonymous referee for pointing us to this connection.}
\begin{thm}
\label{Theorem : main}Suppose Assumptions \ref{Assumption: prior affinity}
and \ref{Assumption: compact convex priors} hold, let the true state
distribution $\mu$ be any admissible prior, and let $\alpha$ be
any action. The following three statements are equivalent:
\begin{enumerate}
\item[(i)] There exists an information structure $(\Sigma,E)$ that implements
$\alpha$ under true state distribution $\mu$.
\item[(ii)] There exists an admissible prior $\nu\in\mathcal{P}$ such that
\begin{enumerate}
\item[(a)] $U(\alpha,\mu)\geq U(\alpha,\nu)\geq U(\beta,\nu)$ for all actions
$\beta$; and
\item[(b)] if $U(\alpha,\mu)>U(\alpha,\nu)$ then $\mu+\lambda(\nu-\mu)\notin\mathcal{P}$
for every $\lambda>1$.
\end{enumerate}
\item[(iii)] There exists an almost fully informative information structure $(\Sigma,E)$
that implements $\alpha$ under true state distribution $\mu$.
\end{enumerate}
\end{thm}
Theorem \ref{Theorem : main} does two things. First, we completely
characterize implementability via geometric augmentations to (\ref{display: saddle inequalities}).
Second, and in keeping with Theorem \ref{thm Manipulability}, we
confirm that implementability is equivalent to implementability via
almost fully informative information structures. As a final matter,
we return to the researcher's problem. In light of Theorem \ref{Theorem : main},
this problem is equivalent to maximizing $v$ on the set of implementable
actions.

\begin{prop}
\label{prop: researcher existence}Suppose Assumptions \ref{Assumption: prior affinity}
and \ref{Assumption: compact convex priors} hold. If the true state
distribution $\mu$ belongs to the relative interior of $\mathcal{P}$
then the set of actions implementable under $\mu$ is non-empty and
compact. Consequently, there exists an almost fully informative information
structure $(\Sigma,E)$ and an action $\alpha$ such that $(\alpha,\Sigma,E)$
solves the researcher's problem.
\end{prop}
Theorem \ref{Theorem : main} provides the substance of Proposition
\ref{prop: researcher existence}, in which our additional hypothesis
about the location of $\mu$ in the relative interior of $\mathcal{P}$
serves to ensure that the set of implementable actions is well behaved.

\section{\label{section: conclusion}Discussion and conclusions}

This paper proposes a model of \emph{identification design }in which
the decision maker observes the entire population-level distribution
of messages generated by an information structure, views any admissible
state distribution that is consistent with the distribution of those
messages as plausible, and ranks actions by their worst-case payoff
over the set of all such distributions. We provide a general characterization
of\emph{ manipulable }environments in which there exists an information
structure that implements each action under each true distribution
of the state of the world. In manipulable environments, an almost
fully informative information structure implements the researcher's
preferred action and solves her problem under every state distribution
in the resulting identified set. We apply our framework to robust
causal inference in econometrics. Finally, we provide a general characterization
of implementability for non-manipulable environments.

Worst-case payoff maximization is one of many ambiguity-averse preferences
considered in the decision theory and statistical decision theory
literatures. We make two comments. First, in keeping with the nonlinearity
of the decision maker's underlying payoff functional 
\[
U(a,\nu)=\mathbb{E}_{\nu}[Y_{a}]=\sum_{y,x}\frac{y}{\nu(a\mid x)}\nu(y,a,x)
\]
in treatment-effects models with uncertainty about the assignment
mechanism, we allow for but do not require the inner payoff function
$U$ to be of the expected utility form. This lies in contrast to
both the maxmin expected utility criterion (\citealp{GilboaSchmeidler1989})
and the standard formulation of its counterparts, including the smooth
ambiguity aversion (\citealp{KlibanoffMarinacciMukerji2005}) and
variational preferences models (\citealp{MaccheroniMarinacciRustichini2006})
that nest maxmin expected utility either directly or as a limiting
case.

Second, regardless of whether or not $U$ represents an expected utility
preference, the outer minimization operation over the identified set
is consistent with the major premise of this paper in the sense that
its only inputs are (i) the structural restrictions to the set of
admissible distributions $\mathcal{P}$ and (ii) the restrictions
on $\mathcal{P}$ implied by the information revealed by the information
structure $(\Sigma,E)$. By way of contrast, both of the aforementioned
alternative models take either second-order beliefs (the former) or
penalty terms (the latter) as exogenous inputs that prioritize some
admissible priors over others. 

Nevertheless, maxmin is not the only suitable criterion for purely
data-driven analysis. In Appendix \ref{AppendX: regret}, we show
that our manipulability result holds under the alternative minimax
regret criterion for a decision maker who restricts attention to pure-strategy
treatments, but does not hold without such a restriction. While the
payoff-guarantee interpretation of maxmin is more economically straightforward
than the psychological interpretation of the regret criterion, this
is a matter of taste. More broadly, unlike results that follow from
the monotonicity of the decision maker's payoff in the size of the
uncertainty set (\citealp{Rosenthal2026PFB}), our results do not
immediately extend to general ambiguity-averse preferences. 

Beyond a richer treatment of alternative ambiguity attitudes, we suggest
several avenues for future work. First, extensions of our framework
to dynamic settings or to games with multiple players. Second, a treatment
of the finite-sample issues that we have deliberately abstracted from
in order to maintain our focus on identification. Third, applications
to ordinary least squares, instrumental variables, and other standard
models in econometrics other than the potential-outcomes setting.
More broadly, we view our model as a starting point for the analysis
of strategic information disclosure to data-driven decision makers.

\appendix

\section{\label{appendix: manip}Technical material, Section \ref{Section: Manip}}
\begin{proof}[Proof of Proposition \ref{Prop which null}]
Let $(\Sigma,E)$ be any information structure and note $\ker E$
is a linear subspace of $\mathbb{R}^{\vert\Omega\vert}$ because $E$
is linear. For each $d\in\ker E$, we have
\[
\sum_{\omega\in\Omega}d(\omega)=\sum_{\omega\in\Omega}d(\omega)\sum_{\sigma\in\Sigma}E(\sigma\mid\omega)=\sum_{\sigma\in\Sigma}(Ed)(\sigma)=\sum_{\sigma\in\Sigma}0=0,
\]
where the first equality follows from $\sum_{\sigma}E(\sigma\mid\omega)=1$
for each state $\omega$ and the third from $d\in\ker E$. 

Conversely, let $D$ satisfy the stated conditions and note $\text{dim}(D)\leq\vert\Omega\vert-1$
because $D$ is a subset of the $\vert\Omega\vert-1$ dimensional
hyperplane $H\equiv\{d\in\mathbb{R}^{\vert\Omega\vert}\mid\sum_{\omega}d(\omega)=0\}$.
There are two trivial cases. First, if $\dim(D)=\vert\Omega\vert-1$
then $D=H$ and thus any information structure with $\Sigma\equiv\{\sigma\}$
and $E(\sigma\mid\omega)\equiv1$ for all states $\omega$ satisfies
$\ker E=D$. Second, if $\text{dim}(D)=0$ then $D=\{0\}$ and any
fully informative information structure $(\Sigma,E)$ satisfies $\ker E=D$. 

Suppose instead $0<\text{dim}(D)<\vert\Omega\vert-1$ and write $n\equiv\vert\Omega\vert$.
We proceed constructively. Let $W$ be the orthogonal complement of
$D$, note $\ell\equiv\text{dim}(W)=n-\dim(D)$, and let $w^{1},...,w^{\ell}$
be a basis for $W$. For each $1\leq i\leq\ell$, let scalars $x^{i},y^{i}$
satisfy $x^{i}>-\min(w^{i})$ and $y^{i}>\max(w^{i})$. Define 
\begin{align*}
p^{i} & \equiv(x^{i},...,x^{i})+w^{i}, & q^{i} & \equiv(y^{i},...,y^{i})-w^{i}, & \lambda & \equiv\frac{1}{\sum_{i}(x^{i}+y^{i})}.
\end{align*}
Let $M$ be the $m\times n$ real matrix with $m\equiv2\ell$ rows
each defined by
\[
(M_{i1},...,M_{in})\equiv\begin{cases}
\lambda p^{i} & i\leq\ell\\
\lambda q^{i-\ell} & i>\ell.
\end{cases}
\]
Finally, let $\Sigma$ be any set of messages with $m$ distinct elements,
enumerate $\Omega$ by $\omega_{1},...,\omega_{n}$, and define experiment
$E$ by $E(\cdot\mid\omega_{j})\equiv(M_{1j},...,M_{mj})$ for each
column $j$ of $M$. Because the entries of $M$ are non-negative
per our choice of $p^{i},q^{i}$ and
\begin{align*}
\sum_{i=1}^{m}M_{ij} & =\sum_{i=1}^{\ell}\lambda(x^{i}+y^{i})=1,
\end{align*}
$(\Sigma,E)$ is a valid information structure. We claim $\ker E=D$.
First, because $w^{1},...,w^{\ell}$ is a basis for $W$ and 
\[
w^{i}=\frac{y^{i}p^{i}-x^{i}q^{i}}{x^{i}+y^{i}}
\]
for each $i$, the set $W$ is contained in the row space of $M$.
Second, because $(1,...,1)\in W$ per our hypothesis $\sum_{\omega}d(\omega)=0$
for all $d\in D$ and $w^{i}\in W$ for all $i$ by definition, we
have $(M_{i1},...,M_{in})\in W$ for all $i$. Accordingly, the row
space of $M$ is contained in $W$. Finally, because we have just
established that the row space of $M$ is the orthogonal complement
of $D$, the fundamental theorem of linear algebra implies $\ker E=D$.
\end{proof}
\begin{proof}[Proof of Theorem \ref{thm Manipulability}]
We proceed by arguing first that (i) implies (ii) and second that
(ii) implies (iii); that (iii) implies (i) is definitional. For the
purposes of establishing the first step by contraposition, suppose
environment $(\Omega,\mathcal{P},A,U)$ and actions $\alpha,\beta$
are such that 
\[
\inf_{\nu\in\mathcal{P}}\;U(\beta,\nu)>\inf_{\nu\in\mathcal{P}}U(\alpha,\nu).
\]
Let $\mu$ be any true state distribution satisfying $\inf_{\nu\in\mathcal{P}}U(\beta,\nu)>U(\alpha,\mu)$
and let $(\Sigma,E)$ be any information structure. Per our choice
of $\mu$, we have 
\[
\inf_{\nu\in\mathcal{P}_{\mu}(\Sigma,E)}U(\alpha,\nu)\leq U(\alpha,\mu)<\inf_{\nu\in\mathcal{P}}U(\beta,\nu)\leq\inf_{\nu\in\mathcal{P}_{\mu}(\Sigma,E)}U(\beta,\nu).
\]
Accordingly, $(\Sigma,E)$ does not implement action $\alpha$ and
$(\Omega,\mathcal{P},A,U)$ is not manipulable.

For the purposes of establishing the second step directly, suppose
(ii) holds and let $\beta$ be any action with full support. First,
by (ii), $\inf_{\nu}U(a,\nu)=\inf_{\nu}U(\beta,\nu)$ for all $a$.
Second, because the image of $\mathcal{P}$ under $U(\beta,\cdot)$
is compact by assumption, there exists an admissible prior $\nu^{*}$
such that $U(\beta,\nu^{*})=\inf_{\nu}U(\beta,\nu)$. Third, the linearity
of the decision maker's payoffs in actions implies
\[
U(\beta,\nu^{*})=\int_{A}U(a,\nu^{*})\D\beta(a).
\]
Because $U$ is continuous and $\beta$ has full support, we conclude
$\inf_{\nu}U(a,\nu)=U(a,\nu^{*})$ for all $a$. In turn, affinity
of the decision maker's utility in actions implies the same equality
holds for every mixed action. Finally, let the true state distribution
$\mu$ be admissible. Proposition \ref{Prop which null} provides
an information structure $(\Sigma,E)$ with $\ker E=\text{span}\{\mu-\nu^{*}\}$.
Any such information structure is almost fully informative and implements
every action. Accordingly, (iii) holds. 
\end{proof}

\section{\label{appendix: te manip}Technical material, Section \ref{Section: TE}}
\begin{proof}[Proof of Proposition \ref{prop: manip}]
Let $(\Omega,\mathcal{P},A,U)$ be any treatment-effects model. We
establish strong manipulability via Theorem \ref{thm Manipulability}.
Let $P$ be any admissible assignment mechanism, choose group weight
$\lambda\in\Delta(\mathcal{X})$ with full support, and define prior
\[
\nu^{*}(y,t,x)\equiv\mathbf{1}\{y=\min\mathcal{Y}\}P(t\mid x)\lambda(x).
\]
First, because
\[
\nu^{*}(t\mid x)=\frac{\nu^{*}(t,x)}{\nu^{*}(x)}=\frac{P(t\mid x)\lambda(x)}{\sum_{t'}P(t'\mid x)\lambda(x)}=P(t\mid x)
\]
for all $(t,x)$, $\nu^{*}$ is admissible. Second, because
\begin{align*}
U(a,\nu) & =\sum_{y,x}y\nu(y\mid a,x)\nu(x)\geq\min\mathcal{Y}=\sum_{y,x}y\nu^{*}(y\mid a,x)\nu^{*}(x)=U(a,\nu^{*})
\end{align*}
for all $a$ and for all admissible $\nu$, we also have 
\begin{equation}
U(\alpha,\nu)=\sum_{a}U(a,\nu)\alpha(a)\geq\sum_{a}U(a,\nu^{*})\alpha(a)=\min\mathcal{Y}=U(\alpha,\nu^{*})\label{display: immiseration}
\end{equation}
for all $\alpha$ and all admissible $\nu$. Accordingly, $\min_{\nu\in\mathcal{P}}U(\alpha,\nu)=\min\mathcal{Y}$
for all actions $\alpha$. Theorem \ref{thm Manipulability} then
implies $(\Omega,\mathcal{P},A,U)$ is strongly manipulable. 
\end{proof}

\section{\label{appendix: non manip}Technical material, Section \ref{sec:Non-manipulable-environments}}

\begin{lem}
\label{Lemma: maxmin to saddle}Suppose Assumptions \ref{Assumption: prior affinity}
and \ref{Assumption: compact convex priors} hold. Action $\alpha$
solves the decision maker's problem under information structure $(\Sigma,E)$
and true state distribution $\mu$ if and only if there exists $\nu\in\mathcal{P_{\mu}}(\Sigma,E)$
such that $(\alpha,\nu)$ is a saddle point of the decision maker's
problem.
\end{lem}
\begin{proof}
Let $(\Sigma,E)$ be any information structure and $\mu$ any admissible
true distribution of the state variable. Suppose $\alpha$ solves
the decision maker's problem under $(\Sigma,E)$ and $\mu$. Because
$\mathcal{P}$ is compact and convex and the map $\nu\mapsto E\nu$
is linear, the identified set $\mathcal{P}_{\mu}(\Sigma,E)$ is compact
and convex. Further, because $A$ is a compact metric space, $\Delta(A)$
is compact and convex. In turn, because $U$ is continuous and affine
in each argument, Sion's minimax theorem implies the decision maker's
problem has a saddle point $(\alpha^{*},\nu^{*})$. Let $\nu$ be
a minimizer for continuous function $U(\alpha,\cdot)$ on compact
set $\mathcal{P_{\mu}}(\Sigma,E)$ and note
\[
U(\alpha,\nu)=\max_{\alpha'\in\Delta(A)}\;\min_{\nu'\in\mathcal{P_{\mu}}(\Sigma,E)}\;U(\alpha',\nu')=\min_{\nu'\in\mathcal{P_{\mu}}(\Sigma,E)}\;\max_{\alpha'\in\Delta(A)}\;U(\alpha',\nu')=U(\alpha^{*},\nu^{*})
\]
because $\alpha$ is a solution and $(\alpha^{*},\nu^{*})$ is a saddle
point. First, because $\nu$ is a minimizer for $U(\alpha,\cdot)$;
$U(\alpha,\nu)=U(\alpha^{*},\nu^{*})$; and $\alpha^{*}$ is a maximizer
for $U(\cdot,\nu^{*})$, we have
\[
\forall\alpha'\in\Delta(A)\;U(\alpha,\nu^{*})\geq U(\alpha,\nu)=U(\alpha^{*},\nu^{*})\geq U(\alpha',\nu^{*}).
\]
Second, because $\alpha^{*}$ is a maximizer for $U(\cdot,\nu^{*})$;
$U(\alpha,\nu)=U(\alpha^{*},\nu^{*})$; and $\nu$ is a minimizer
for $U(\alpha,\cdot)$, we have
\[
\forall\nu'\in\mathcal{P_{\mu}}(\Sigma,E)\;U(\alpha,\nu^{*})\leq U(\alpha^{*},\nu^{*})=U(\alpha,\nu)\leq U(\alpha,\nu').
\]
Accordingly, $(\alpha,\nu^{*})$ is a saddle point, as claimed. Conversely,
if there exists $\nu\in\mathcal{P}_{\mu}(\Sigma,E)$ such that $(\alpha,\nu)$
is a saddle point of the decision maker's problem, then the saddle
inequalities imply immediately that $\alpha$ is a solution to that
problem.
\end{proof}
\Needspace{4\baselineskip}
\begin{proof}[Proof of Theorem \ref{Theorem : main}]
As in the Proof of Theorem \ref{thm Manipulability}, we proceed
by showing first (i) implies (ii) and second (ii) implies (iii); that
(iii) implies (i) is again definitional. Proceeding with the first
step, suppose $(\Sigma,E)$ implements $\alpha$. By Lemma \ref{Lemma: maxmin to saddle},
there exists an admissible prior $\nu$ satisfying $E\nu=E\mu$ such
that $(\alpha,\nu)$ is a saddle point of the decision maker's problem.
It follows immediately that for every action $\beta$
\[
U(\alpha,\mu)\geq\min_{\nu'\in\mathcal{P}_{\mu}(\Sigma,E)}\;U(\alpha,\nu')=U(\alpha,\nu)\geq U(\beta,\nu)
\]
and thus (ii)(a) holds. If $U(\alpha,\nu)=U(\alpha,\mu)$ then (ii)(b)
is satisfied by definition. Otherwise, if $U(\alpha,\nu)<U(\alpha,\mu)$
then any admissible prior $\mu+\lambda(\nu-\mu)$ with $\lambda>1$
belongs to $\mathcal{P}_{\mu}(\Sigma,E)$ and satisfies $U(\alpha,\mu+\lambda(\nu-\mu))<U(\alpha,\nu)$,
contrary to our conclusion that $(\alpha,\nu)$ is a saddle point.
Accordingly, (ii)(b) holds. In both cases, (ii) holds. 

Proceeding with the second step, suppose admissible prior $\nu$ satisfies
(ii) and let information structure $(\Sigma,E)$ satisfy $\ker E=\text{span}\{\nu-\mu\}$,
as justified by Proposition \ref{Prop which null}. For every action
$\beta$ we have
\[
\min_{\nu'\in\mathcal{P}_{\mu}(\Sigma,E)}\;U(\alpha,\nu')=U(\alpha,\nu)\geq U(\beta,\nu)\geq\min_{\nu'\in\mathcal{P}_{\mu}(\Sigma,E)}\;U(\beta,\nu'),
\]
where the equality follows from Assumption \ref{Assumption: prior affinity}
and (ii)(a)--(b), the first inequality follows from (ii)(a), and
the second inequality follows from $\nu\in\mathcal{P}_{\mu}(\Sigma,E)$.
Accordingly, almost fully informative information structure $(\Sigma,E)$
implements $\alpha$. 
\end{proof}
\begin{proof}[Proof of Proposition \ref{prop: researcher existence}]
Let $\Delta^{*}$ denote the set of actions implementable under $\mu$
and let $G\subset\Delta(A)\times\mathcal{P}$ be the set of pairs
$(\alpha,\nu)$ satisfying condition (ii) of Theorem \ref{Theorem : main}.
Because $\mu$ belongs to the relative interior of $\mathcal{P}$,
the set of priors satisfying the termination condition in (ii)(b)
is exactly the relative boundary $B$ of $\mathcal{P}$, which is
closed.

Accordingly, condition (ii) is equivalent to (ii)(a) together with
either $U(\alpha,\mu)=U(\alpha,\nu)$ or $\nu\in B$. Each of these
three conditions defines a closed set of pairs, by continuity of $U$
and closedness of $B$. Hence $G$ is a closed subset of a compact
set and thus itself compact. By Theorem \ref{Theorem : main}, $\Delta^{*}$
is the projection of $G$ onto its first coordinate and is therefore
compact. In turn, because $v$ is upper semicontinuous, it attains
a maximum on $\Delta^{*}$. Theorem \ref{Theorem : main} supplies
an almost fully informative information structure implementing that
action.
\end{proof}

\section{\label{AppendX: regret}Regret minimization}

Consider the \emph{regret minimization proble}m
\[
\min_{\alpha\in\Delta(A)}\;\sup_{\nu\in\mathcal{P_{\mu}}(\Sigma,E)}\;[\max_{\beta\in\Delta(A)}\;U(\beta,\nu)-U(\alpha,\nu)]
\]
in which the decision maker seeks to minimize the worst-case difference
between the full-information value of his decision problem and his
actual value. We develop two results for this decision making criterion.
First, we show that the \emph{regret minimization problem with pure
strategie}s
\[
\min_{a\in A}\;\sup_{\nu\in\mathcal{P_{\mu}}(\Sigma,E)}\;[\max_{b\in A}\;U(b,\nu)-U(a,\nu)]
\]
satisfies a version of our manipulability criterion. Second, we show
that the regret minimization problem itself does not.
\begin{defn}
Environment $(\Omega,\mathcal{P},A,U)$ is \emph{pure-strategy} \emph{regret-manipulable
}if for every true distribution $\mu$ and every action $a\in A$
there exists an information structure $(\Sigma,E)$ such that $a$
is a solution to the regret minimization problem with pure strategies.
\end{defn}
\begin{prop}
\label{prop 4: ps regret}Treatment-effects models $(\Omega,\mathcal{P},A,U)$
are pure-strategy regret-manipulable. 
\end{prop}
\begin{proof}
Let true distribution $\mu$ be any admissible prior. For each $a\in A$
define prior
\[
\nu^{a}(y,t,x)\equiv\begin{cases}
\mathbf{1}\{y=\min\mathcal{Y}\}\mu(t\mid x)\mu(x) & t=a\\
\mathbf{1}\{y=\max\mathcal{Y}\}\mu(t\mid x)\mu(x) & t\neq a.
\end{cases}
\]
First, because $\nu^{a}(t\mid x)=\mu(t\mid x)$ for all $(t,x)$,
$\nu^{a}$ is admissible. Second, Proposition \ref{Prop which null}
provides an information structure $(\Sigma,E)$ with $\ker E=\{d\mid\sum_{\omega}d(\omega)=0\}$.
Because $\nu^{a}\in\mathcal{P}_{\mu}(\Sigma,E)$ for all $a\in A$
and $U(b,\nu^{a})-U(a,\nu^{a})=\max\mathcal{Y}-\min\mathcal{Y}$ for
all $a\in A$ and all $b\neq a$, every action $a\in A$ has maximum
regret $\max\mathcal{Y}-\min\mathcal{Y}$ and thus every action $a\in A$
is a solution to the regret minimization problem with pure strategies.
\end{proof}
\begin{defn}
Environment $(\Omega,\mathcal{P},A,U)$ is \emph{regret-manipulable
}if for every true distribution $\mu$ and every action $\alpha\in\Delta(A)$
there exists an information structure $(\Sigma,E)$ such that $\alpha$
is a solution to the regret minimization problem.
\end{defn}
\begin{prop}
\label{prop 5: regret manip}Treatment-effects models $(\Omega,\mathcal{P},A,U)$
are not regret-manipulable. 
\end{prop}
\begin{proof}
Choose $a\in A$, let $P$ be any admissible assignment mechanism,
and let $\lambda\in\Delta(\mathcal{X})$ be any group weight with
full support. Consider true state distribution
\[
\mu(y,t,x)\equiv\begin{cases}
\mathbf{1}\{y=\min\mathcal{Y}\}P(t\mid x)\lambda(x) & t=a\\
\mathbf{1}\{y=\max\mathcal{Y}\}P(t\mid x)\lambda(x) & t\neq a
\end{cases}
\]
and note that $\mu$ is admissible because $\mu(t\mid x)=P(t\mid x)$
for all $(t,x)$. Let $(\Sigma,E)$ be any information structure,
let $\nu$ be any admissible state distribution, and let $\alpha\in\Delta(A)$
be uniformly distributed. First, because $\mu\in\mathcal{P}_{\mu}(\Sigma,E)$,
$a$ has maximum regret $\max\mathcal{Y}-\min\mathcal{Y}$. Second,
set $M\equiv\max_{\beta\in\Delta(A)}U(\beta,\nu)$ and note that because
(i) $M$ is attained at some pure action $b^{*}\in A$ and (ii) $U(b,\nu)\geq\min\mathcal{Y}$
for all $b\in A$, the regret for action $\alpha$ satisfies
\begin{align*}
M-U(\alpha,\nu) & =M-\frac{1}{\vert A\vert}\sum_{b}U(b,\nu)\leq M-\frac{1}{\vert A\vert}\left(M+(\vert A\vert-1)\min\mathcal{Y}\right)\\
 & =\frac{\vert A\vert-1}{\vert A\vert}(M-\min\mathcal{Y})\leq\frac{\vert A\vert-1}{\vert A\vert}(\max\mathcal{Y}-\min\mathcal{Y}).
\end{align*}
Accordingly, the maximum regret for $\alpha$ is bounded strictly
below $\max\mathcal{Y}-\min\mathcal{Y}$. We conclude $a$ is not
a solution to the regret minimization problem under information structure
$(\Sigma,E)$ and true state distribution $\mu$.
\end{proof}
Propositions \ref{prop 4: ps regret} and \ref{prop 5: regret manip}
establish that treatment-effects environments with regret-minimizing
decision makers are manipulable if and only if the decision maker
restricts attention to pure strategies. While this restriction can
be strictly costly for the decision maker, it is with precedent not
only in robust mechanism design theory\footnote{See \citet{Carroll2015}'s influential robust contracting paper, in
which randomization is ruled out as a modeling choice even though
it is strictly beneficial to the principal in typical specifications
of the problem, as shown in \citet{Kambhampati2023} and \citet{KambhampatiEtAl2025}.} but also in practice, to the extent that real-world policymakers
do not typically randomize over policies.

\end{document}